\documentstyle[prd,aps,preprint,draft]{revtex}
\renewcommand{\(}{\left(}
\renewcommand{\)}{\right)}
\renewcommand{\to}{\rightarrow}
\newcommand{\as}{\alpha_s}
\newcommand{\Ei}{{\rm Ei}}
\newcommand{\kv}{{\bf k}}
\newcommand{\sv}{{\bf S}}
\newcommand{\pv}{{\bf p}}

\newcommand{\gev}{{\rm\ GeV}}
\begin{document}
\thispagestyle{empty}
\draft
\title{Quantum-Chromodynamic Potential Model for Light-Heavy Quarkonia
	and the Heavy Quark Effective Theory}
\author{Suraj N. Gupta and James M. Johnson}
\address{ Department of Physics, Wayne State University, Detroit,
	Michigan 48202}
%\date{\today}
\maketitle
\begin{abstract}\baselineskip=22pt
	We have investigated the spectra of light-heavy quarkonia with the
use of a quantum-chromodynamic potential model
which is similar to that used earlier
for the heavy quarkonia.  An essential feature of our treatment is the
inclusion of the one-loop radiative
corrections to the  quark-antiquark potential,
which contribute significantly to the spin-splittings
among the quarkonium energy levels.
Unlike $c\bar{c}$ and $b\bar{b}$, the potential for a
light-heavy system has a complicated dependence on the light and heavy
quark masses $m$ and $M$, and
it contains a spin-orbit mixing term.  We have
obtained excellent results for the observed energy levels
of $D^0$, $D_s$, $B^0$, and $B_s$, and we are able to
provide predicted results for many unobserved
energy levels.  Our potential parameters for different
quarkonia satisfy the constraints of quantum chromodynamics.

	We have also used our investigation to test the accuracy
of the heavy quark effective theory.  We find that
the heavy quark expansion yields generally good results for the $B^0$
and $B_s$ energy levels provided that $M^{-1}$ and $M^{-1}\ln M$
corrections are taken into account in the quark-antiquark interactions.
It does not, however, provide equally good results for the energy
levels of $D^0$ and $D_s$, which indicates that the effective theory can
be applied more accurately to the $b$ quark than the $c$ quark.
\end{abstract}
\narrowtext

\section{INTRODUCTION}

	The light-heavy quarkonia $D$, $D_s$, $B$, and $B_s$ are
of much experimental and theoretical interest, and their exploration
is necessary for our understanding of the strong as well as the
electroweak interactions\cite{review,theor}.  We shall here investigate
the spectra of light-heavy quarkonia with the use
of a quantum chromodynamic model similar to the highly successful
model used earlier for the heavy
quarkonia $c\bar{c}$ and $b\bar{b}$\cite{ccbar}.  The complexity
of the model is necessarily enhanced for a light-heavy system because
the potential has a complicated dependence on the light
and heavy quark masses $m$ and $M$, and it contains a spin-orbit
mixing term.  We shall obtain results for the observed
and the unobserved energy levels of $D^0(c\bar{u})$, $D_s(c\bar{s})$,
$B^0(b\bar{d})$, and $B_s(b\bar{s})$, compare
them with the available experimental data, and examine their scaling behavior.
The $u$-$d$ mass difference and the electromagnetic interaction will be ignored
in
the present investigation\cite{uord}.

	We shall also use our results to test
the accuracy of  the heavy quark effective theory\cite{iswise,hqet}
both in the limit of
$M\rightarrow\infty$ as well as with the inclusion of the $M^{-1}$
and $M^{-1}\ln M$ corrections.

	The approximate heavy quark symmetry, like the approximate
chiral symmetry, points to an underlying difference
between the $(u,d,s)$ and the $(c,b,t)$ quarks.  It is interesting
that this fundamental difference was recognized in our
mass-matrix approach to quark mixing and CP violation\cite{mmatrix}, which
predicted the value $M_t\leq 170$~GeV for the top quark mass in
excellent agreement with the recently reported experimental value of
$174\pm 17$~GeV\cite{top}.

\section{LIGHT-HEAVY QUARKONIUM SPECTRA}

	Our treatment for the light-heavy quarkonia is similar to that
for $c\bar{c}$ and $b\bar{b}$ \cite{ccbar} except for the complications arising
from the
difference in the quark and antiquark masses.  Thus, our model
is based on the Hamiltonian
\begin{equation}
H=H_0+V_p+V_c,
\end{equation}
where
\begin{equation}
H_0=(m^2+\pv^2)^{1/2}+(M^2+\pv^2)^{1/2}
\end{equation}
is the relativistic kinetic energy term, and $V_p$ and $V_c$ are nonsingular
quasistatic perturbative and confining potentials, which are given in
Appendix~A.  Our trial wave function for obtaining the quarkonium
energy levels and wave functions is of the same form as in
the earlier investigations.  Since our
potentials are nonsingular, we are able to avoid the use of an illegitimate
perturbative treatment.

	The experimental and theoretical results for the energy levels
of the light-heavy quarkonia $D^0$, $B^0$, $D_s$, and $B_s$, together
with the ${}^3P^\prime_1$-${}^1P^\prime_1$ mixing angles
arising from the spin-orbit mixing terms, are given in Tables~I-IV.
For experimental data we have relied on the Particle Data Group\cite{pdg}
except that we have used
the more recent results from the CLEO collaboration\cite{cleo1,cleo10}
for $D_1^0$, $D_2^{\star0}$, and $D_{s2}$ and from the
CDF collaboration\cite{cdf} for $B_s$.
In these tables, one set of theoretical results corresponds to the
direct use of our model, while the other two sets are obtained
by means of heavy quark expansions of our potentials
to test the accuracy of the heavy quark effective theory
with the inclusion of the $M^{-1}$ and $M^{-1}\ln M$ corrections as
well as without these corrections.  The approximate potentials
corresponding to the effective theory are given in Appendix~B.

	We expect the dynamics of a light-heavy system
to be primarily dependent on the light quark.  Therefore, our potential
parameters for $D^0$ and $B^0$ are the same except for the
difference in the $c$ and $b$ quark masses, and they are given by
\begin{eqnarray}
m_{u,d}&=& 0.350\gev,\nonumber\\
M_c&=& 1.690\gev,\nonumber\\
M_b&=& 5.400\gev,\nonumber\\
\mu&=& 0.932\gev,\label{pardb}\\
\as&=& 0.3965,\nonumber\\
A&=& 0.185\gev^2,\nonumber\\
B&=& 0.152.\nonumber
\end{eqnarray}
Similarly, the parameters for $D_s$ and $B_s$ are
\begin{eqnarray}
m_s&=& 0.514\gev,\nonumber\\
M_c&=& 1.578\gev,\nonumber\\
M_b&=& 5.040\gev,\nonumber\\
\mu&=& 1.250\gev,\label{pardsbs}\\
\as&=& 0.340,\nonumber\\
A&=& 0.198\gev^2,\nonumber\\
B&=& 0.131.\nonumber
\end{eqnarray}

	We have ensured that the values of $\alpha_s$, $M_c$, and $M_b$
in (\ref{pardb}) and (\ref{pardsbs}) are related through the quantum
chromodynamic transformation relations
\begin{equation}
\as^\prime=\frac{\as}{1+\beta_0(\as/4\pi) \ln ({\mu^\prime}^2/\mu^2)}\ ,
\end{equation}
and
\begin{equation}
M^\prime=M\(\frac{\as^\prime}{\as}\)^{2\gamma_0/\beta_0},
\end{equation}
where $\beta_0=11-\frac{2}{3}n_f$, $n_f=3$, and $\gamma_0=2$.  The use of the
one-loop transformation relations is consistent with the
inclusion of the one-loop radiative corrections in the quarkonium
potentials. Moreover, since $u$, $d$, and $s$ are the dynamical quarks
in the light-heavy systems, a higher value of $\mu$ for quarkonia with
the $s$ quark is to be expected.

	A precise determination of the potential parameters for the light-heavy
quarkonia is difficult because of the availability of only
limited experimental data.  This difficulty, however, has been mitigated
in our treatment by requiring that the parameters for the four systems
satisfy reasonable physical and quantum-chromodynamic constraints.

	We have also looked at the correlation of our parameters for
the light-heavy quarkonia with those of other quarkonia.  When applied
to $u\bar{d}$ and $u\bar{s}$, our parameters yield good
results for the $\pi$-$\rho$ and $K$-$K^\star$ splittings.  We are,
however, unable to correlate our parameters for the light-heavy quarkonia
with those for the heavy quarkonia through the transformation relations
(5) and (6), and keeping in mind the past success of our quarkonium model
we can only offer the following possible explanation:

	Strictly speaking, the QCD transformation relations are applicable
only to the current quarks.  According to our experience, the transformation
relations seem to hold reasonably well for the heavy quarkonia $c\bar{c}$
and $b\bar{b}$ as well as for the quarkonia containing one or two
light quarks, but they are unsuitable for correlating the parameters
for these two classes of quarkonia.  This seems to be another
manifestation of the difference between light and heavy quarks.
We believe a full explanation would require an understanding of the origin of
the constituent quark masses, which remains unclear at this time.

	Our phenomenological confining potential for the light-heavy quarkonia
is of the same form as that for the heavy quarkonia.  We find that the
parameter $A$ for the spin-independent term in the confining potential
is approximately the same for all quarkonia, while the spin-dependent
terms vary such that the vector-exchange component is smaller for
$c\bar{c}$ than for $b\bar{b}$, and still smaller for the light-heavy
quarkonia.

\section{CONCLUSION}

	We have obtained excellent results for the observed energy levels
of $D^0$, $B^0$, $D_s$, and $B_s$ with the use of our quantum-chromodynamic
potential model, and provided predicted
results for many unobserved energy levels in Tables I-IV.
We have included in these tables the mixing angles for the $1\;{}^3P^\prime_1$
and $1\;{}^1P^\prime_1$ levels, which are needed for an understanding
of their decay properties.
Although the use of a semirelativistic model may seem questionable for a system
containing a light quark, ultimately such an approach should be judged on
the basis of its predictions\cite{uddif}.
Additional experimental data on the light-heavy quarkonia should
be available in the near future.

	We have also used our results to test the accuracy of the heavy quark
effective theory\cite{iswise,hqet}.  By comparing the theoretical
results without and with the effective theory in Tables~I-IV, we find
that the heavy quark
expansion with the inclusion of the $M^{-1}$ and $M^{-1}\ln M$ corrections
yields generally good results for the $B^0$ and $B_s$ energy levels.
It does not, however, provide equally good results for the
energy levels of $D^0$ and $D_s$, which indicates that the effective theory
can be applied more accurately to the $b$ quark than the
$c$ quark\cite{randall}.

	We further find that the results for the energy levels in the
limit $M\to\infty$ are unacceptable.  As is well known, in this limit
the energy level pairs $({}^3S_1,{}^1S_0)$, $({}^3P^\prime_1,{}^3P_0)$,
and $({}^3P_2,{}^1P^\prime_1)$ become degenerate.

	Finally, we have examined the scaling behavior of energy level splittings
in the light-heavy quarkonia by looking at the results obtained by the
direct use of our model in Tables I-IV.  As shown in Tables V and VI,
the splittings between levels which become degenerate in the limit
$M\to\infty$ exhibit an approximate $M^{-1}$ scaling.  This scaling
behavior does not apply to splittings between other
pairs of energy levels.

\acknowledgments
        This work was supported in part by the U.S. Department of Energy
under Grant No.~DE-FG02-85ER40209.

\appendix
\section{NONSINGULAR QUARKONIUM POTENTIALS}

	The nonsingular potentials for a light-heavy
quarkonium are similar to those used recently for $c\bar{c}$ \cite{ccbar}
except for the complications due to the difference in the quark and
antiquark masses.  The complications are further enhanced by the conversion
of the singular potentials\cite{GR} into the nonsingular
ones\cite{SNG}, which are
necessary to avoid the use of an illegitimate perturbative
treatment.  The corresponding denominators in the
singular and nonsingular potentials for a quark and an antiquark
of masses $m_1$ and $m_2$ are related as
\begin{equation}
\frac{1}{m_1^2}\to\frac{1}{m_1^2+\frac{1}{4}\kv^2}\ ,\qquad
	\frac{1}{m_2^2}\to\frac{1}{m_2^2+\frac{1}{4}\kv^2}\ ,
\end{equation}
and
\[
\frac{1}{m_1m_2}\to\frac{1}{(m_1^2+\frac{1}{4}\kv^2)^{1/2}
	(m_2^2+\frac{1}{4}\kv^2)^{1/2}}\approx
	\frac{1}{m_1m_2+\frac{1}{8}\(\frac{m_1}{m_2}
		+\frac{m_2}{m_1}\)\kv^2}
\]
or
\begin{equation}
\frac{1}{m_1m_2}\to\frac{\kappa}{\omega^2+\frac{1}{4}\kv^2}\ ,
\end{equation}
where
\begin{equation}
\kappa=\frac{2m_1m_2}{m_1^2+m_2^2}\ ,\qquad\omega=(\kappa m_1m_2)^{1/2}\;.
\end{equation}

	The potentials for a quark and an antiquark of
different flavors are given below. They reduce to those for
$c\bar{c}$ for $m_1=m_2=m$ except that, unlike $c\bar{c}$, they
do not contain the annihilation terms.

\widetext
\subsection{Perturbative quantum-chromodynamic potential}

	The potential in the momentum space is
\begin{eqnarray}
%		Spin-Independent
V_p(\kv)&=& -\frac{16\pi\as}{3\kv^2}  \left[1-\frac{3\as}{2\pi}
		-\frac{\as}{12\pi}(33-2n_f)
		\ln\(\frac{\kv^2}{\mu^2}\)\right]\nonumber\\
& &\hspace*{0.75in}+\frac{8\pi\as}{3} \(\frac{1}{\kv^2+4m_1^2}
	+\frac{1}{\kv^2+4m_2^2}\)\left[ \delta_{l0}\(1-\frac{3\as}{2\pi}\)
	-\frac{\as}{12\pi} (33-2n_f) \ln\(\frac{\kv^2}{\mu^2}\)
		\right]\nonumber\\
& & \hspace*{0.75in}-\frac{8\pi^2\as^2 }{9|\kv|}
	\frac{\kappa}{\kv^2+4\omega^2}
		\left[9(m_1+m_2)-\frac{8m_1m_2}{m_1+m_2}\right]\nonumber\\[8pt]
%
%		Hyperfine
& & \hspace*{0.in}+\frac{64\pi\as}{3}\;
		\sv_1\!\cdot\!\sv_2\frac{\kappa}{\kv^2+4\omega^2}
	\left\{\delta_{l0}\left[\frac{2}{3}-\frac{19\as}{9\pi}
	-\frac{\as}{12\pi}\(8\frac{m_1-m_2}{m_1+m_2}+\frac{m_1+m_2}{m_1-m_2}\)
		\ln\(\frac{m_2}{m_1}\)\right]\right.\nonumber\\
& & \hspace*{0.75in}\left.-\frac{\as}{18\pi}(33-2n_f)\ln\(\frac{\kv^2}{\mu^2}\)
	+\frac{7\as}{4\pi}\ln\(\frac{\kv^2}{m_1m_2}\)
	\right\}\nonumber\\[8pt]
%
%		Tensor
& & \hspace*{0.in}-\frac{64\pi\as}{3}\;
	\frac{\sv_1\!\cdot\kv\ \sv_2\!\cdot\!\kv
	-\frac{1}{3}\kv^2\sv_1\!\cdot\!\sv_2}{\kv^2}
		\frac{\kappa}{\kv^2+4\omega^2}\left[1+\frac{4\as}{3\pi}
	-\frac{\as}{12\pi} (33-2n_f)
		\ln\(\frac{\kv^2}{\mu^2}\)\right.\nonumber\\
& & \hspace*{0.75in}\left. +\frac{3\as}{2\pi}\ln\(\frac{\kv^2}{m_1m_2}\)\right]
		\nonumber\\[8pt]
%
%		Spin-Orbit
& & \hspace*{0.in} -\frac{16\pi\as}{3}\; \frac{i\sv\cdot(\kv\times\pv)}{\kv^2}
	\left\{\(\frac{1}{\kv^2+4m_1^2}+\frac{1}{\kv^2+4m_2^2}\)
		\left[1-\frac{\as}{6\pi}
	-\frac{\as}{12\pi} (33-2n_f) \ln\(\frac{\kv^2}{\mu^2}\)
	\right.\right]\nonumber\\
& & \hspace*{0.75in}+\frac{3\as}{2\pi}\frac{1}{\kv^2+4m_1^2}
		\ln\(\frac{\kv^2}{m_1^2}\)
	+\frac{3\as}{2\pi} \frac{1}{\kv^2+4m_2^2}
		\ln\(\frac{\kv^2}{m_2^2}\)\nonumber\\
& & \hspace*{0.75in}\left.+\frac{4\kappa}{\kv^2+4\omega^2}
	\left[1-\frac{5\as}{6\pi}
	-\frac{\as}{12\pi} (33-2n_f) \ln\(\frac{\kv^2}{\mu^2}\)
	+\frac{3\as}{4\pi} \ln\(\frac{\kv^2}{m_1m_2}\) \right]
		\right\}\nonumber\\[8pt]
%
%		Mixing
& & \hspace*{0.in}-\frac{16\pi\as}{3} \;
	\frac{i(\sv_1-\sv_2)\cdot(\kv\times\pv)}{\kv^2}
		\left\{\(\frac{1}{\kv^2+4m_1^2}-\frac{1}{\kv^2+4m_2^2}\)
		\left[1-\frac{\as}{6\pi}
	-\frac{\as}{12\pi} (33-2n_f) \ln\(\frac{\kv^2}{\mu^2}\)
	\right.\right]\nonumber\\
& & \hspace*{0.75in}\left.+\frac{3\as}{2\pi}\frac{1}{\kv^2+4m_1^2}
		\ln\(\frac{\kv^2}{m_1^2}\)
	-\frac{3\as}{2\pi}\frac{1}{\kv^2+4m_2^2}\ln\(\frac{\kv^2}{m_2^2}\)
		+\frac{3\as}{2\pi}\frac{2\kappa}{\kv^2+4\omega^2}
		\ln\(\frac{m_2}{m_1}\)\right\}.
\end{eqnarray}
\smallskip

	In the coordinate space, it takes the form
\smallskip
\begin{eqnarray}
%	Spin-Indep.
V_p({\bf r})&=& -\frac{4\as}{3r}\left\{1-\frac{3\as}{2\pi}+\frac{\as}{6\pi}
	(33-2n_f)\left[\ln(\mu r)+\gamma_E\right]
	 -\delta_{l0}\left(1-\frac{3\as}{2\pi}\right)\left(
	\frac{e^{-2m_1r}+e^{-2m_2r}}{2}\right)\right. \nonumber \\
& &\hspace*{0.75in}-\frac{\as}{12\pi}(33-2n_f)\left[\ln(\mu r)
		\left(e^{-2m_1r}+e^{-2m_2r}\right)
		+E_+(2m_1r)+E_+(2m_2r)\right]\nonumber\\
& &\hspace*{0.75in}+\left. \frac{\as}{6}\left[
		9(m_1+m_2)-\frac{8m_1m_2}{m_1+m_2}\right]\frac{\kappa}{\omega}
		\left[\ln(2\omega r)e^{-2\omega r}-E_-(2\omega r)\right]
		\right\}\nonumber\\[8pt]
%
%	Hyperfine
& & +\kappa\frac{32\as}{9r}\sv_1\cdot\sv_2\left\{ \delta_{l0}\left[1-
	\frac{19\as}{6\pi}-\frac{\as}{8\pi}\left(
	8\frac{m_1-m_2}{m_1+m_2}+\frac{m_1+m_2}{m_1-m_2}\right)
		\ln\left(\frac{m_2}{m_1}\right)
		\right]e^{-2\omega r}\right.\nonumber\\
& &\hspace*{0.75in}\left. +\frac{\as}{6\pi}(33-2n_f)
	\left[\ln(\mu r)e^{-2\omega r}+E_+(2\omega r)\right]
	-\frac{21\as}{4\pi}\left[\ln(\sqrt{m_1m_2}\;r)e^{-2\omega r}
		+E_+(2\omega r)\right]\right\}\nonumber\\[8pt]
%
%	Tensor
& & +\kappa\frac{4\as}{3r}S_T\left\{ \(1+\frac{4\as}{3\pi}\) f_2(2\omega r)
	+\frac{\as}{6\pi}(33-2n_f)\left[f_2(2\omega r)\ln(\mu r)
		+g_2(2\omega r)\right]\right.\nonumber\\
& & \hspace*{0.75in}\left. -\frac{3\as}{\pi}\left[f_2(2\omega r)
		\ln(\sqrt{m_1m_2}\;r)+g_2(2\omega r)\right]\right\}
		\nonumber\\[8pt]
%
%	Spin-Orbit
& & +\frac{4\as}{3r}{\bf L}\cdot\sv\left\{\(1-\frac{\as}{6\pi}\)
		\left[f_1(2m_1r)+f_1(2m_2r)\right]
	\right.\nonumber\\
& & \hspace*{0.75in}+\frac{\as}{6\pi}(33-2n_f)\left[f_1(2m_1r)\ln(\mu r)
	+g_1(2m_1r)+f_1(2m_2r)\ln(\mu r)+g_1(2m_2r)\right]\nonumber\\
& & \hspace*{0.75in}-\frac{3\as}{\pi}\left[f_1(2m_1r)\ln(m_1 r)
	+g_1(2m_1r)+f_1(2m_2r)\ln(m_2 r)+g_1(2m_2r)\right]\nonumber\\
& & \hspace*{0.75in}+4\kappa\left[\(1-\frac{5\as}{6\pi}\) f_1(2\omega r)
	+\frac{\as}{6\pi}(33-2n_f)\left[f_1(2\omega r)
	\ln(\mu r)+g_1(2\omega r)\right]\right.\nonumber\\
& & \hspace*{1.0in}\left.\left.-\frac{3\as}{2\pi}\left[f_1(2\omega r)
	\ln(\sqrt{m_1m_2}\;r)+g_1(2\omega r)\right]\right]
\right\}\nonumber\\[8pt]
%
%	Mixing
& &+\frac{4\as}{3r}{\bf L}\cdot(\sv_1-\sv_2)\left\{ \(1-\frac{\as}{6\pi}\)
	\left[f_1(2m_1r)-f_1(2m_2r)\right]\right.\nonumber\\
& &\hspace*{0.75in}+\frac{\as}{6\pi}(33-2n_f)\left[f_1(2m_1r)\ln(\mu r)
	+g_1(2m_1r)-f_1(2m_2r)\ln(\mu r)-g_1(2m_2r)\right]\nonumber\\
& &\hspace*{0.75in}-\frac{3\as}{\pi}\left[f_1(2m_1r)\ln(m_1r)
	+g_1(2m_1r)-f_1(2m_2r)\ln(m_2 r)-g_1(2m_2r)\right]\nonumber\\
& &\hspace*{0.75in}\left.+\frac{3\as}{\pi}\ln\left(\frac{m_2}{m_1}\right)
	\kappa f_1(2\omega r)\right\}\ .
\end{eqnarray}

	Note that the tensor operator is defined as
\begin{equation}
S_T=3\ \mbox{\boldmath$\sigma$}_1\cdot{\bf \hat{r}}\
	\mbox{\boldmath$\sigma$}_2\cdot{\bf \hat{r}}
	-\mbox{\boldmath$\sigma$}_1\cdot\mbox{\boldmath$\sigma$}_2,
\end{equation}
the functions $E_\pm$ are expressible in terms of the
exponential-integral function $\Ei$ as
\begin{equation}
E_\pm(x)= \frac{1}{2}\left[e^x \Ei(-x)\pm e^{-x}\Ei(x)\right]
	\mp e^{-x}\ln x,
\end{equation}
and
\begin{eqnarray}
f_1(x)&=& \frac{1-(1+x)e^{-x}}{x^2},\nonumber\\[2pt]
f_2(x)&=& \frac{1-\left(1+x+\frac{1}{3}x^2\right)e^{-x}}{x^2},\nonumber\\[2pt]
g_1(x)&=& \frac{\gamma_E-\left[E_+(x)-xE_-(x)\right]}{x^2},\label{defns}\\[2pt]
g_2(x)&=& \frac{\gamma_E-\left[\left(1+\frac{1}{3}x^2\right)E_+(x)
	-xE_-(x)\right]}{x^2}.\nonumber
\end{eqnarray}

\subsection{Phenomenological confining potential}

	The scalar-vector-exchange confining potential is given by
\begin{equation}
V_c=(1-B)V_S+BV_V,
\end{equation}
where in the momentum space
\begin{eqnarray}
V_S({\bf k})&=& -8\pi A\left[\frac{1}{\kv^4}
	-\frac{i\sv\cdot(\kv\times\pv)}{\kv^4}
	\(\frac{1}{\kv^2+4m_1^2}+\frac{1}{\kv^2+4m_2^2}\)\right.\nonumber\\
& &\hspace{0.75in}\left.-\frac{i(\sv_1-\sv_2)\cdot(\kv\times\pv)
		}{\kv^4}\(\frac{1}{\kv^2+4m_1^2}-\frac{1}{\kv^2+4m_2^2}\)
		\right] ,
\end{eqnarray}
\begin{eqnarray}
V_V({\bf k})&=&  -8\pi A\left[\frac{1}{{\bf k}^4}
	-\frac{1}{2\kv^2}\(\frac{1}{\kv^2+4m_1^2}+\frac{1}{\kv^2+4m_2^2}\)
	-\frac{8\kappa\sv_1\cdot\sv_2}{3\kv^2(\kv^2+4\omega^2)}
	+\frac{4\kappa\(\sv_1\!\cdot\kv\ \sv_2\!\cdot\!\kv
	-\frac{1}{3}\kv^2\sv_1\!\cdot\!\sv_{2}\)}{\kv^4
		(\kv^2+4\omega^2)}\right.\nonumber\\
& &\hspace{0.75in}+\frac{i\sv\cdot(\kv\times\pv)}{\kv^4}
		\(\frac{1}{\kv^2+4m_1^2}+\frac{1}{\kv^2+4m_2^2}
			+\frac{4\kappa}{\kv^2+4\omega^2}\)\nonumber\\
& &\hspace{0.75in}\left. +\frac{i(\sv_1-\sv_2)\cdot(\kv\times\pv)}{\kv^4}
		\(\frac{1}{\kv^2+4m_1^2}-\frac{1}{\kv^2+4m_2^2}\)\right] .
\end{eqnarray}

	The coordinate-space potentials are
\begin{eqnarray}
V_S({\bf r})&=& Ar-\frac{A}{4r}{\bf L}\cdot\sv\left[
	\frac{1-2f_1(2m_1r)}{m_1^2}+\frac{1-2f_1(2m_2r)}{m_2^2}\right] \nonumber\\
& &\hspace*{0.75in}-\frac{A}{4r} {\bf L}\cdot(\sv_1-\sv_2)\left[
	\frac{1-2f_1(2m_1r)}{m_1^2}-\frac{1-2f_1(2m_2r)}{m_2^2}\right]
		\ ,\\[12pt]
V_V({\bf r})&=& Ar+\frac{A}{4r}\left(\frac{1-e^{-2m_1r}}{m_1^2}
	+\frac{1-e^{-2m_2r}}{m_2^2}\right)
	+\frac{4A}{3r} \sv_1\cdot\sv_2
		\left(\frac{1-e^{-2\omega r}}{m_1m_2}\right)
	+\frac{A}{12r}S_T\left[\frac{1-6f_2(2\omega r)}{m_1m_2}\right]
		\nonumber\\
& & \hspace*{0.75in}+\frac{A}{r}{\bf L}\cdot\sv\left[
	\frac{1-2f_1(2m_1r)}{4m_1^2}+\frac{1-2f_1(2m_2r)}{4m_2^2}
	+\frac{1-2f_1(2\omega r)}{m_1m_2}\right]\nonumber\\
& &\hspace*{0.75in}+\frac{A}{4r} {\bf L}\cdot(\sv_1-\sv_2)\left[
	\frac{1-2f_1(2m_1r)}{m_1^2}-\frac{1-2f_1(2m_2r)}{m_2^2}\right]
		\ .
\end{eqnarray}

	It is understood that the confining potential also contains an additive
phenomenological constant $C$.

\section{QUARKONIUM POTENTIALS WITH HEAVY QUARK EXPANSION}

	Upon replacing $m_1$ and $m_2$ by $m$ and $M$, and expanding in powers
of $M^{-1}$, the coordinate-space potentials of Appendix~A take the
approximate forms given below.

\subsection{Perturbative Potential}
\begin{equation}
V_p({\bf r})=V_{p0}({\bf r})+\left(\frac{m}{M}\right) V_{p1}({\bf r})
	+{\cal O}\bigg(\frac{m^2}{M^2}\bigg)
\end{equation}
with
\begin{eqnarray}
V_{p0}({\bf r})&=& -\frac{4\as}{3r}\left\{1-\frac{3\as}{2\pi}+\frac{\as}{6\pi}
	(33-2n_f)\left[\ln(\mu r)+\gamma_E\right]
	 -\delta_{l0}\left(1-\frac{3\as}{2\pi}\right)
	\frac{e^{-2mr}}{2}\right. \nonumber \\
& &\hspace*{0.75in}-\frac{\as}{12\pi}(33-2n_f)\left[\ln(\mu r)e^{-2mr}
		+E_+(2mr)\right]\nonumber\\
& &\hspace*{0.75in}+\left. \frac{3\as}{\sqrt{2}}
		\left[\ln(2\sqrt{2} mr)e^{-2\sqrt{2}m r}-E_-(2\sqrt{2}m r)\right]
		\right\}\nonumber\\[8pt]
%
%	Spin-Orbit
& & +\frac{4\as}{3r}{\bf L}\cdot\sv\left\{\(1-\frac{\as}{6\pi}\)f_1(2mr)
	+\frac{\as}{6\pi}(33-2n_f)\left[f_1(2mr)\ln(\mu r)
	+g_1(2mr)\right]\right.\nonumber\\
& & \hspace*{0.75in}\left.-\frac{3\as}{\pi}\left[f_1(2mr)\ln(m r)
	+g_1(2mr)\right]\right\}\nonumber\\[8pt]
%
%	Mixing
& &+\frac{4\as}{3r}{\bf L}\cdot(\sv_1-\sv_2)\left\{ \(1-\frac{\as}{6\pi}\)
	f_1(2mr)+\frac{\as}{6\pi}(33-2n_f)\left[f_1(2mr)\ln(\mu r)
	+g_1(2mr)\right]\right.\nonumber\\
& &\hspace*{0.75in}\left.-\frac{3\as}{\pi}\left[f_1(2mr)\ln(mr)
	+g_1(2mr)\right]\right\},\label{vp0}\\[12pt]
V_{p1}({\bf r})&=& -\frac{2\sqrt{2}\as^2}{9r}
	\left[\ln(2\sqrt{2}m r)e^{-2\sqrt{2}m r}-E_-(2\sqrt{2}m r)\right]
		\nonumber\\[8pt]
%
%	Hyperfine
& & +\frac{64\as}{9r}\sv_1\cdot\sv_2\left\{ \delta_{l0}\left[1-
	\frac{19\as}{6\pi}+\frac{9\as}{8\pi}
		\ln\left(\frac{M}{m}\right)
		\right]e^{-2\sqrt{2}m r} \right.\nonumber\\
& &\hspace*{0.75in}\left.+\frac{\as}{6\pi}(33-2n_f)
	\left[\ln(\mu r)e^{-2\sqrt{2}m r}+E_+(2\sqrt{2}m r)\right]
		\right.\nonumber\\
& &\hspace*{0.75in}\left.
	-\frac{21\as}{4\pi}\left[\ln(\sqrt{mM}\;r)e^{-2\sqrt{2}m r}
		+E_+(2\sqrt{2}m r)\right]\right\}\nonumber\\[8pt]
%
%	Tensor
& & +\frac{8\as}{3r}S_T\left\{ \(1+\frac{4\as}{3\pi}\) f_2(2\sqrt{2}m r)
	+\frac{\as}{6\pi}(33-2n_f)\left[f_2(2\sqrt{2}m r)\ln(\mu r)
		+g_2(2\sqrt{2}m r)\right]\right.\nonumber\\
& & \hspace*{0.75in}\left. -\frac{3\as}{\pi}\left[f_2(2\sqrt{2}m r)
	\ln(\sqrt{mM}\;r)+g_2(2\sqrt{2}m r)\right]\right\}\nonumber\\[8pt]
%
%	Spin-Orbit
& & +\frac{32\as}{3r}{\bf L}\cdot\sv\left\{
	\(1-\frac{5\as}{6\pi}\) f_1(2\sqrt{2}m r)
	+\frac{\as}{6\pi}(33-2n_f)\left[f_1(2\sqrt{2}m r)
	\ln(\mu r)+g_1(2\sqrt{2}m r)\right]\right.\nonumber\\
& & \hspace*{0.75in}\left.-\frac{3\as}{2\pi}\left[f_1(2\sqrt{2}m r)
	\ln(\sqrt{mM}\;r)+g_1(2\sqrt{2}m r)\right]\right\}\nonumber\\[8pt]
%
%	Mixing
& &+\frac{8\as^2}{\pi r}{\bf L}\cdot(\sv_1-\sv_2)
	\ln\left(\frac{M}{m}\right) f_1(2\sqrt{2}m r)\ .
\end{eqnarray}

\subsection{Confining potentials}
\begin{equation}
V_S({\bf r})=V_{S0}({\bf r})+{\cal O}\bigg(\frac{m^2}{M^2}\bigg)
\end{equation}
with
\begin{equation}
V_{S0}({\bf r})=Ar-\frac{A}{4r}{\bf L}\cdot\sv
	\frac{1-2f_1(2mr)}{m^2} -\frac{A}{4r} {\bf L}\cdot(\sv_1-\sv_2)
	\frac{1-2f_1(2mr)}{m^2}	\ ,\label{vs0}
\end{equation}
and
\begin{equation}
V_V({\bf r})=V_{V0}({\bf r})+\left(\frac{m}{M}\right)V_{V1}({\bf r})
	+{\cal O}\bigg(\frac{m^2}{M^2}\bigg)
\end{equation}
with
\begin{eqnarray}
V_{V0}({\bf r})&=& Ar+\frac{A}{4r}\;\frac{1-e^{-2mr}}{m^2}
	+\frac{A}{4r}{\bf L}\cdot\sv
	\frac{1-2f_1(2mr)}{m^2}+\frac{A}{4r} {\bf L}\cdot(\sv_1-\sv_2)
	\frac{1-2f_1(2mr)}{m^2}\ ,\label{vv0}\\[18pt]
V_{V1}({\bf r})&=& \frac{4A}{3r} \sv_1\cdot\sv_2
		\frac{1-e^{-2\sqrt{2}m r}}{m^2}
	+\frac{A}{12r}S_T\frac{1-6f_2(2\sqrt{2}m r)}{m^2}
	+\frac{A}{r}{\bf L}\cdot\sv\frac{1-2f_1(2\sqrt{2}m r)}{m^2}\ .
\end{eqnarray}

\narrowtext
\subsection{$M\to\infty$ limit}

	In the limit $M\to\infty$, the perturbative and confining
potentials are given by
\begin{equation}
V_p=V_{p0},\qquad V_S=V_{S0},\qquad V_V=V_{V0}.
\end{equation}
According to (\ref{vp0}), (\ref{vs0}), and (\ref{vv0}), the
spin-orbit and spin-orbit-mixing terms in these potentials are
of the form
\begin{equation}
f(r){\bf L}\cdot \sv + f(r){\bf L}\cdot(\sv_1-\sv_2),\label{mixed}
\end{equation}
which is expressible solely in terms of the light-quark spin
as
\begin{equation}
2f(r){\bf L}\cdot\sv_1 .
\end{equation}
We have, however, given the potentials in the form (\ref{mixed})
to facilitate comparison with the more accurate treatments of
the light-heavy quarkonia.

\narrowtext

\begin{table}
\caption{\label{d0} $D^0$ energy levels
in MeV.  Effective theory results are given with the $M^{-1}$ and $M^{-1}\ln M$
corrections as well as in the limit of $M\to\infty$.
Experimental results are from Refs.~\protect\cite{pdg}
and \protect\cite{cleo1}. }
\bigskip
\begin{tabular}{lr@{$\pm$}lccc}
\hspace*{1pt}&\multicolumn{2}{c}{Expt.}&Theory&Effective theory&$M\to\infty$\\
\tableline
$1\;{}^1S_0\,\ (D^0)$&	1864.5&0.5	&	1864.5& 1864.5& 1864.5\\
$1\;{}^3S_1\,\ (D^{\star 0})$&2007&1.4	&	2007.0& 2010.9& 1864.5\\
$2\;{}^1S_0\, $&	\multicolumn{2}{c}{}	&	2547.7& 2566.5& 2431.9\\
$2\;{}^3S_1\, $&	\multicolumn{2}{c}{}	&	2647.0& 2662.1& 2431.9\\
$1\;{}^3P_0\, $&  \multicolumn{2}{c}{}	&	2278.6& 2310.2& 2244.8\\
$1\;{}^3P^\prime_1\, $& \multicolumn{2}{c}{} &	2407.3& 2414.6& 2244.8\\
$1\;{}^3P_2\,\ (D^0_2)$&	2465&4.2	&	2465.0& 2474.0& 2287.2\\
$1\;{}^1P^\prime_1\,\ (D^0_1)$&2421&2.8	&	2421.0& 2438.2& 2287.2\\
$\theta$&	\multicolumn{2}{c}{}	&	29.0$^\circ$&	30.9$^\circ$&	35.6$^\circ$
\end{tabular}
\end{table}

\begin{table}
\caption{\label{ds} $D_s$ energy levels
in MeV. Experimental results are from Refs.~\protect\cite{pdg}
and \protect\cite{cleo10}. }
\bigskip
\begin{tabular}{lr@{$\pm$}lccc}
\hspace*{1pt}&\multicolumn{2}{c}{Expt.}&Theory&Effective theory&$M\to\infty$\\
\tableline
$1\;{}^1S_0\,\ (D_s)$&	1968.8&0.7	&	1968.8& 1968.8& 1968.8\\
$1\;{}^3S_1\,\ (D^\star_s)$&2110.3&2.0	&	2110.5& 2113.1& 1968.8\\
$2\;{}^1S_0\, $&	\multicolumn{2}{c}{}	&	2656.5& 2678.8& 2536.5\\
$2\;{}^3S_1\, $&	\multicolumn{2}{c}{}	&	2757.8& 2774.3& 2536.5\\
$1\;{}^3P_0\, $&  \multicolumn{2}{c}{}	&	2387.8& 2422.2& 2382.2\\
$1\;{}^3P^\prime_1\, $& \multicolumn{2}{c}{} &	2521.2& 2528.8& 2382.2\\
$1\;{}^3P_2\,\ (D_{s2})$&	2573.2&1.9	&	2573.1& 2582.8& 2402.8\\
$1\;{}^1P^\prime_1\,\ (D_{s1})$&2536.5&0.8	&	2536.5& 2552.1& 2402.8\\
$\theta$&	\multicolumn{2}{c}{}	&	26.0$^\circ$&	31.8$^\circ$&	35.6$^\circ$
\end{tabular}
\end{table}

\begin{table}
\caption{\label{b0} $B^0$ energy levels
in MeV.  Experimental results are from Ref.~\protect\cite{pdg}. }
\bigskip
\begin{tabular}{lr@{$\pm$}lccc}
\hspace*{1pt}&\multicolumn{2}{c}{Expt.}&Theory&Effective theory&$M\to\infty$\\
\tableline
$1\;{}^1S_0\,\ (B^0)$&	5278.7&2.1	&	5278.7& 5278.7& 5278.7\\
$1\;{}^3S_1\,\ (B^{\star 0})$&5324.6&2.1	&	5324.0& 5325.8& 5278.7\\
$2\;{}^1S_0\, $&	\multicolumn{2}{c}{}	&	5892.1& 5893.9& 5846.3\\
$2\;{}^3S_1\, $&	\multicolumn{2}{c}{}	&	5924.3& 5927.1& 5846.3\\
$1\;{}^3P_0\, $& \multicolumn{2}{c}{}	&	5689.5& 5692.5& 5659.1\\
$1\;{}^3P^\prime_1\, $& \multicolumn{2}{c}{} &	5730.8& 5734.1& 5659.1\\
$1\;{}^3P_2\, $&	\multicolumn{2}{c}{}	&	5759.1& 5761.4& 5701.5\\
$1\;{}^1P^\prime_1\, $& \multicolumn{2}{c}{}&	5743.6& 5745.4& 5701.5\\
$\theta$&	\multicolumn{2}{c}{}	&	31.7$^\circ$&	31.3$^\circ$&	35.6$^\circ$
\end{tabular}
\end{table}

\begin{table}
\caption{\label{bs} $B_s$ energy levels
in MeV. Experimental results are from Refs.~\protect\cite{pdg}
and \protect\cite{cdf}. }
\bigskip
\begin{tabular}{lr@{$\pm$}lccc}
\hspace*{1pt}&\multicolumn{2}{c}{Expt.}&Theory&Effective theory&$M\to\infty$\\
\tableline
$1\;{}^1S_0\,\ (B_s)$&	5383.3&6.7	&	5383.3& 5383.3& 5383.3\\
$1\;{}^3S_1\,\ (B_s^\star)$&5430.5&2.6	&	5431.9& 5434.1& 5383.3\\
$2\;{}^1S_0\, $&	\multicolumn{2}{c}{}	&	6000.9& 6003.1& 5950.9\\
$2\;{}^3S_1\, $&	\multicolumn{2}{c}{}	&	6035.8& 6039.1& 5950.9\\
$1\;{}^3P_0\, $& \multicolumn{2}{c}{}	&	5810.1& 5814.2& 5796.7\\
$1\;{}^3P^\prime_1\, $& \multicolumn{2}{c}{} &	5855.0& 5857.9& 5796.7\\
$1\;{}^3P_2\, $&	\multicolumn{2}{c}{}	&	5875.2& 5878.1& 5817.1\\
$1\;{}^1P^\prime_1\, $& \multicolumn{2}{c}{}&	5860.2& 5863.2& 5817.1\\
$\theta$&	\multicolumn{2}{c}{}	&	27.3$^\circ$&	27.1$^\circ$&	35.6$^\circ$
\end{tabular}
\end{table}

\begin{table}
\caption{\label{split0} Scaling of energy level splittings in
$D^0$ and $B^0$. Splittings are given in MeV. }
\bigskip
\begin{tabular}{lddd}
 &$\Delta D_0$&$(m_c/m_b)\Delta D^0$&$\Delta B^0$\\
\tableline
$1\;{}^3S_1-1\;{}^1S_0$&	142.5&	44.6&	45.3\\
$2\;{}^3S_1-2\;{}^1S_0$&	99.3&	31.1&	32.2\\
$1\;{}^3P_1^\prime-1\;{}^3P_0$&	128.7&	40.3&	41.3\\
$1\;{}^3P_2-1\;{}^1P_1^\prime$&	44.0&	13.8&	15.5
\end{tabular}
\end{table}

\begin{table}
\caption{\label{splits} Scaling of energy level splittings in
$D_s$ and $B_s$. Splittings are given in MeV. }
\bigskip
\begin{tabular}{lddd}
 &$\Delta D_s$&$(m_c/m_b)\Delta D_s$&$\Delta B_s$\\
\tableline
$1\;{}^3S_1-1\;{}^1S_0$&	141.7&	44.4&	48.6\\
$2\;{}^3S_1-2\;{}^1S_0$&	101.3&	31.7&	35.0\\
$1\;{}^3P_1^\prime-1\;{}^3P_0$&	133.4&	41.8&	44.9\\
$1\;{}^3P_2-1\;{}^1P_1^\prime$&	36.6&	11.5&	15.0
\end{tabular}
\end{table}


\begin{thebibliography}{99}
\bibitem{uord} Although our treatment is equally applicable
	to quarkonia containing either a $u$ or a $d$~quark, we do
	not have as much experimental data on $D^\pm$ and $B^\pm$ as we
	have on $D^0$ and $B^0$.
\bibitem{uddif}Our predicted
	values for the energy levels do not include the
	effects of the $u$-$d$ mass difference and the electromagnetic
	interaction, which may amount to a few MeV.
\bibitem{randall} The accuracy with which the effective theory can
	be applied to $c$ quarks has also been questioned by
	L.~Randall and E.~Sather, Phys. Rev. D {\bf 49}, 6236 (1994).
\end{thebibliography}
\end{document}